# Mirror-coupled plasmonic bound states in the continuum for tunable perfect absorption

Juan Wang[1], Thomas Weber[1], Andreas Aigner[1], Stefan A. Maier[2,3,1] and Andreas Tittl[1]*

[1]Chair in Hybrid Nanosystems, Nanoinstitute Munich, Faculty of Physics, Ludwig-Maximilians-Universität München, München 80539, Germany

[2]School of Physics and Astronomy, Monash University, Clayton, Victoria 3800, Australia

[3]The Blackett Laboratory, Department of Physics, Imperial College London, London SW7 2AZ, United Kingdom

*e-mail: andreas.tittl@physik.uni-muenchen.de

**Abstract**

Tailoring critical light-matter coupling is a fundamental challenge of nanophotonics, impacting diverse fields from higher harmonic generation and energy conversion to surface-enhanced spectroscopy. Plasmonic perfect absorbers (PAs), where resonant antennas couple to their mirror images in adjacent metal films, have been instrumental for obtaining different coupling regimes by tuning the antenna-film distance. However, for on-chip uses, the ideal PA gap size can only match one wavelength, and wide range multispectral approaches remain challenging. Here, we introduce a new paradigm for plasmonic PAs by combining mirror-coupled resonances with the unique loss engineering capabilities of plasmonic bound states in the continuum (BICs). Our BIC-driven PA platform leverages the asymmetry of the constituent meta-atoms as an additional degree of freedom for reaching the critical coupling (CC) condition, delivering resonances with unity absorbance and high quality factors approaching 100 in the mid-infrared. Such a platform holds flexible tuning knobs including asymmetry parameter, dielectric gap, and geometrical scaling factor to precisely control the coupling condition, resonance frequency, and selective enhancement of magnetic and electric fields while maintaining CC. We demonstrate a pixelated PA metasurface with optimal absorption over a broad range of mid-infrared frequencies (950 ~ 2000 cm$^{-1}$) using only a single spacer layer thickness and apply it for multispectral surface-enhanced molecular spectroscopy in tailored coupling regimes. Our concept greatly expands the capabilities and flexibility of traditional gap-tuned PAs, opening new perspectives for miniaturized sensing platforms towards on-chip and in-situ detection.

**Introduction**

Metamaterials have attracted intensive attention in many fields such as holography,[1–3] energy conversion,[4–6] lasers,[7,8] biomedical imaging and diagnosis,[9–11] lightweight augmented reality and virtual reality headsets and others,[12–14] due to their capabilities of control over light in miniaturized optical devices with tunable resonances, programmable phases and enhanced near fields. One landmark metamaterial concept for tailoring light-matter coupling is the plasmonic perfect absorber (PA), where the precise interference between a resonant antenna and its own mirror image in a neighboring metal layer separated by a dielectric spacer is leveraged to provide tunable resonances and near-unity absorbance. These advantages have led to the broad adoption of PAs for diverse applications including sensitive detectors,[15] energy harvesting,[6,16,17] thermal emitters,[18] and molecular sensing.[19–23] In general, plasmonic PAs are affected by the intrinsic losses inherent to their metallic components, limiting the quality ($Q$) factors of resonances, defined as the resonance position divided by the full width at half maximum (FWHM) intensity. Recently, the use of magnetic Mie resonances supported by low-loss dielectric materials coupled to metallic films has been proposed to produce resonances with high $Q$ factors.[24] However, such magnetic resonances confine electromagnetic energy (and therefore the electromagnetic near fields) mostly inside of the all-dielectric resonators, leading to significant drawbacks for the enhancement of surface-driven coupling processes like sensing or photocatalysis, where the target system is placed adjacent to the metasurface.[25] Consequently, it remains challenging to engineer a PA geometry which simultaneously supports high $Q$ factors, strongly enhanced and surface-confined near fields, and precisely controlled light-matter coupling. In particular, practical point-of-care spectroscopy devices in biomedical assays require optimal analyte sensitivity and high specificity over a broad wavelength range to identify and quantify target molecules. Towards this goal, different configurations of PAs were designed and demonstrated for sensing applications.[21,26,27] The main design principles of metallic PAs include surface lattice resonances (SLRs),[26] gap-induced plasmons,[23] and metal-insulator-metal (MIM) stacks.[28,29] PAs based on SLRs[26,30,31] and vertical gap induced plasmons[23,32] have achieved sharp resonances with $Q$ factors of typically 100 in the visible spectral region, but tuning the resonances over a wide spectral range with near-unity absorbance is still challenging. To date, there are only a few reports of metallic PAs with sharp resonances in the mid-infrared (mid-IR) regime, even though they are crucial for surface-enhanced infrared absorbance spectroscopy (SEIRAS) to fingerprint molecules with high specificity compared to refractometric sensing in the visible regime.[33,34] Similarly, PAs with MIM configurations have been widely explored for near-unity absorbance from the visible to the microwave regime,[28,35–37] but the $Q$ factors of such resonant systems are generally low on the order of 10. In addition, to identify different molecular species with spectrally separated characteristic vibrational bands, PAs based on MIM structures lack the flexibility to tailor multiple resonances over a broad IR spectral range with near-unity absorbance and high $Q$ factors simultaneously. Dual-resonant PAs for molecular fingerprinting have been demonstrated using symmetric[21] and asymmetric crossed bars,[33] as well as single disks[38], but still struggle to resolve vibrational bands over a broad spectral range, which is essential to, e.g., resolve complex mixtures of molecules.

To obtain critical light-matter coupling with plasmonic PAs based on MIM structures, controlling and matching the radiative loss channel to the intrinsic loss of the system is crucial, which is mainly implemented through the thickness of the insulator layer.[33,39,40] Harnessing the insulator thickness as the only tuning parameter for controlling the radiative loss makes it extremely difficult to maintain optimum coupling, unity absorbance, and high $Q$ resonances over a broad spectral range. To overcome this limitation, a new degree of freedom for controlling the radiative loss channel is needed. Photonic BICs have recently been introduced as a new class of resonant states that cannot couple to the far field and can therefore exhibit infinite $Q$ factors and enhanced electromagnetic near fields.[41–43] In practice, small parameter changes (geometric or excitation conditions) can perturb the BIC, leading to the emergence of quasi BICs that are accessible from the far-field. The two leading approaches for the realization of quasi BICs are symmetry-protected BIC metasurfaces[44,45] and accidental BICs created by the destructive interference of multiple resonant modes.[7]

Here, we demonstrate a new concept for engineering the optical response of metallic PAs in the mid-IR spectral range by leveraging a MIM structure composed of a resonant plasmonic BIC metasurface, a dielectric spacer layer with constant thickness, and a metallic mirror. Our BIC-driven PA scheme provides two degrees of freedom to control the radiative loss channel, where both the thickness of the insulator layer and asymmetry parameter of the meta-atoms can be used to tune the losses in the system. Significantly, at a given constant gap $g$, the in-plane tilting angle $\theta$ of the meta-atoms can be used to freely vary the radiative loss, allowing to reach the CC conditions under any gaps at will. This two-dimensional parameter space $(g, \theta)$ allows us to freely select between either highly enhanced electric or magnetic fields while maintaining the CC condition or to maximize the value of the total $Q$ factor for given asymmetries. This is a crucial departure from established PA concepts, where the optical response is generally dominated by the choice of a fixed insulator layer thickness. Our design achieves high $Q$ factor up to 100 in the mid-IR, addressing the limitations of inherent metal loss and demonstrating five-fold larger values compared to conventional metallic PAs. Meanwhile, the enhanced near-fields ($|\mathbf{E}|/|\mathbf{E}_0|^2$) of our plasmonic BIC-driven PA approach are significantly larger ($> 10^4$) in comparison with low-loss all-dielectric metasurfaces ($\sim 10^3$),[46] demonstrating strong surface confinement surrounding the meta-atoms. By simultaneously tailoring both the scaling ($S$) factors of the meta-atoms and the asymmetric parameter ($\alpha$, $\alpha = \sin \theta$) with a fixed gap, we further demonstrate a pixelated BIC-driven PA metasurface with discrete ultrasharp resonances in the mid-IR, showing high spectral resolution, molecular specificity, and uniform sensitivity for SEIRAS over a broad spectral range.

## Results

**Principle and numerical design of the plasmonic BIC-driven PA**

The BIC-driven PA consists of a periodic array of tilted ellipse pairs separated from a gold (Au) ground plane by a dielectric spacer layer of calcium fluoride ($CaF_2$) as shown in **Figure 1a**. The polarization of the illuminated incident light is along the short axis of ellipses and the incident light propagation is perpendicular to the metasurfaces. For our subsequent analysis, the crucial structural parameters of this arrangement are the tilting angle $\theta$ of the ellipses, which controls the asymmetry characteristics of the plasmonic BIC mode, and the thickness $g$ of the dielectric layer, which determines the interaction between the plasmonic BIC and induced mirror BIC. To obtain a BIC-driven PA design with the desired resonance wavelength and mode structure, we numerically optimized the geometric parameters of a unit cell including the periodicities in both $x$ and $y$ direction ($P_{x,0}$, $P_{y,0}$), the long/short axis of ellipses ($A_0$, $B_0$), and the height of ellipses ($h$). Once a suitable design for a specific wavelength is obtained, a universal scaling approach is implemented to produce resonances in other spectral regions at will through scaling of the unit cell geometry by a factor $S$ (**Figure S1**, Supporting Information). As mentioned before, conventional PAs based on MIM structures mostly rely on tuning the spacer thickness $g$.[37,40] This established tuning process is maintained in our design; however, our PA concept incorporating plasmonic BICs creates a new degree of freedom for tailoring the absorption enhancement and coupling regime, significantly expanding current capabilities by leveraging the precise control over radiative loss channels provided by the asymmetric parameter $\alpha$ of the metasurface layer (**Figure 1a**).

Our numerical simulations of the current density at resonance (for details see Materials and Methods) show clear plasmonic BIC and induced mirror BIC modes with anti-parallel circulating currents and associated magnetic dipole moments (**Figure 1b**). The current flow in the tilted metallic ellipse pairs shows a characteristic anti-parallel behavior associated with BIC-based unit cells, which was previously observed in the dipolar displacement currents of all-dielectric BIC metasurfaces.[46] The emergence of this distinct mode in the proposed MIM structure indicates that the nature of the BIC mode associated with the tilted ellipse pairs is maintained within the full BIC-driven PA structure. Based on the definition of the absorbance ($A = 1 - R - T$), perfect absorption requires the efficient suppression of both the reflectance ($R$) and transmittance ($T$) of the nanophotonic structure. In our design, $T = 0$ is satisfied by the presence of the optically thick Au mirror (200 nm), which prevents the incident light from penetrating through the structure. The suppression of $R$ is achieved by tailoring the spacer layer thickness ($g$) and the asymmetric parameter ($\theta$), which can be understood within the framework of optical impedance matching[47] or by considering the interference between different layers.[48]

**Figure 1c** shows the absorbance spectra with varying gap ($g$) at a fixed ellipse tilting angle of $\theta = 20º$, demonstrating near-unity absorbance (> 99.9 %) at $g = 200$ nm. Away from this optimum value of $g$, both the maximum absorbance and $Q$ factor of the PA resonance decreases rapidly (**Figure 1d**), indicating that conventional gap tuning still plays a significant role. But importantly, in our BIC-driven PA concept, the magnitude and spectral location of absorption can be controlled even for a fixed value of $g = 200$ nm by tuning the in-plane asymmetric parameter $\alpha$ of the meta-atoms (**Figure 1e**). This approach creates one more degree of freedom to tailor the radiative loss, which constitutes a significant advantage for MIM-based PAs, because the same value of $g$ cannot normally be used to simultaneously realize resonances with both near-unity absorbance and high $Q$ factors over a wide spectra region for real-world applications.[33,38] **Figures 1d** and **1f** show the corresponding absorbance values and $Q$ factors in both $g$- and $\theta$- tailored systems. Notably, we found that the total $Q$ factor in the $(g, \theta)$-tailored system shows a tunable optimal value depending on the interaction between BIC and mirror-BIC in the near field (**Figure S2**). Our approach can enable us to realize multiple PA pixels with widely different resonance wavelengths but uniform $Q$ factors on a single substrate with constant $g$ via geometric scaling by a factor $S$, compared to the established gap-tuned absorber system (**Figure S3**). In the symmetric case ($\theta = 0°$), the additional radiative leakage channel provided by the BIC mode disappears, removing the coupling of PA to the far field and leading to negligible absorbance (**Figure S4a**), which is analogous to the ideal BIC case introduced previously.[49,50] For the symmetric case, PA operation can be recovered by modifying the geometric parameters, but this results in a conventional PA design with much lower $Q$ factor (**Figures S4b** and **c**).

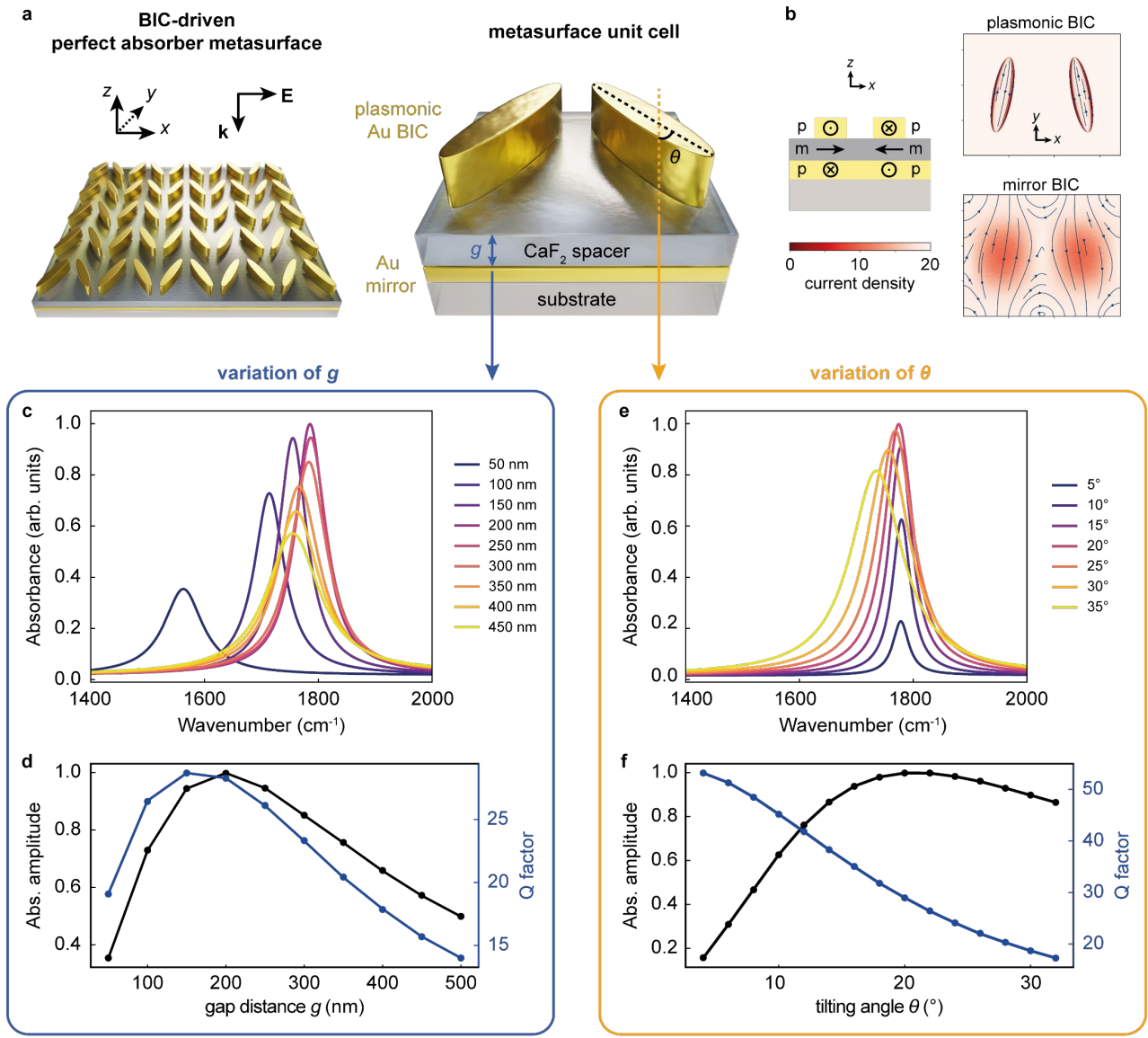

**Figure 1**. Operating principle and schematic of the plasmonic BIC-driven PA concept. (**a**) Metasurface overview and close up diagram of a unit cell of a PA platform consisted of paired tilting ellipses and an Au ground plane separated by a dielectric $CaF_2$ layer. The gap size (*g*) associated with the dielectric spacer and the asymmetric parameter (*θ*) of ellipse are highlighted. (**b**) Side-view of a unit cell indicating the current flow and induced magnetic dipole responses, and current density map of the plasmonic and mirror BICs. (**c**) Absorbance spectra of a BIC-driven PA with constant *θ* = 20° and varying *g*. (**d**) Extracted absorbance amplitudes and *Q* factors for the system with varying *g* at a constant *θ* = 20°. (**e**) Absorbance spectra of a BIC-driven PA with constant *g* = 200 nm and varying *θ*. (**f**) Extracted absorbance amplitudes and *Q* factors for the system with varying *θ* for a gap of g = 200 nm.

**Practical realization of plasmonic BIC-driven PAs**

The tunability of the BIC-driven PA resonance, enabled both by the spacer layer thickness $g$ and the ellipse tilting angle $\theta$, opens up a full two-dimensional parameter space for realizing perfect absorption. Optimal transfer of incident light energy to the PA structure occurs at the so-called CC condition, where the intrinsic $Q$ factor ($Q_{int}$) equals the external (radiative) $Q$ factor ($Q_{rad}$).[40] It is worth noting that the radiative loss rate $\gamma_{rad}$ includes contributions from both the gap size associated with the spacer layer ($\gamma_{rad, g}$) and the asymmetry ($\gamma_{rad, \theta}$) of the plasmonic BIC. We examine two representative examples of CC occurring in disparate areas of the parameter space. **Figures 2a** and **b** compare the electric (**E**) and magnetic (**H**) near-fields of PAs with a thin gap ($g = 100$ nm, and $\theta = 47$ °) and a thick gap ($g = 600$ nm, and $\theta = 8$ °). For the PA design with a thin dielectric gap, the **H** field is strongly enhanced and confined in the dielectric layer, but only a comparatively low enhancement of the **E** field is observed (**Figure 2a**). In contrast, for the PA with a large gap, the situation is reversed, showing weaker **H** field in the gap, but strong **E** field enhancement surrounding the ellipse tips, which agrees well with the previously reported field patterns for dielectric BIC metasurfaces.[46,50] This behavior highlights the unique capability of the BIC-driven PA design to select between **E** and **H** field enhancement while maintaining critical light-matter coupling. Furthermore, the **E** and **H** field vector distributions for both parameter cases clearly show electric dipoles with antiparallel orientation along the long axes of the ellipses as well circulating magnetic field loops surrounding the ellipses with similarly opposing directions (**Figures 2a** and **b**). This observation further confirms that the PA modes maintain their BIC character even for markedly different field enhancement behavior.

To generalize our observations and provide a fundamental understanding of BIC-driven PAs, we map and analyze a large section of the ($g$, $\theta$) parameter space with both numerical simulations (**Figures 2c** and **d**) and experimental measurements (**Figures 2e** and **f**). The simulated absorbance map reveals that for any given spacer layer thickness $g$, there is an associated value of $\theta$ that yields the CC condition (indicated by the dashed black line) and therefore perfect absorption. This behavior is reproduced in the experimental absorption maps, which are in good qualitative agreement with numerical predictions apart from a slight offset in $g$. Notably, larger values of the spacer layer thickness $g$ require smaller tilting angles $\theta$ to achieve the CC condition and vice versa. According to the previous work[37,40] and the following analytical results, we assume that the intrinsic loss of the system $\gamma_{int}$ remains mostly constant for changing spacer layer thickness, we can attribute the observed CC relationship to a rebalancing of the radiative loss channels between $\gamma_{rad, g}$ and $\gamma_{rad, \theta}$. It has been shown previously that $\gamma_{rad, g}$ increases with $g$. Consequently, for larger values of $g$, the radiative loss $\gamma_{rad, g}$ dominates and a BIC unit cell with small $\theta$ and an associated small value of $\gamma_{rad, \theta}$ has to be chosen. The same argument holds for small $g$, where large values of $\gamma_{rad, \theta}$ are required. In addition, these insights can be used to explain why the **H** field is strongly confined in dielectric layer for smaller $g$, which is due to the smaller $\gamma_{rad, g}$ but larger $\gamma_{rad, \theta}$. Conversely, the **E** field is enhanced surrounding the ellipse tips for smaller $\theta$ because of the reduced $\gamma_{rad, \theta}$. Our results indicate that the choice of $g$ is more influential for tailoring the strength of the magnetic mode in the system and $\theta$ is responsible for the electric field confinement governed by the plasmonic

BICs. The CC line subdivides the parameter space into the regimes of under coupling (UC), where the $\gamma_{int} > (\gamma_{rad, g} + \gamma_{rad, \theta})$ and over coupling (OC), where $\gamma_{int} < (\gamma_{rad, g} + \gamma_{rad, \theta})$. The multi-tuning capability of the BIC-driven PA allows to precisely target and realize a specific coupling condition (UC, CC, or OC). Since deviation from the CC condition reduces the **E** and **H** fields enhancements due to the imbalance between $\gamma_{int}$ and $\gamma_{rad}$ (**Figure S5**), parameter regions close to the CC line will be chosen for the later biospectroscopy demonstrations.

**Figures 2c** and **e** show exemplarily simulated and experimental absorbance spectra of BIC-driven PAs with a fixed $g$ = 600 nm (simulations) and $g$ = 580 nm (experiments) for varying $\theta$. The experimental absorbance spectra agree very well with the numerical predictions, but different values of the ellipse tilting angle were required to reach the CC point, with $\theta$ = 8° in simulations, and $\theta$ = 15° in experiments. Such deviations can be explained by additional experimental losses due to the surface roughness of evaporated $CaF_2$ and Au films (see Materials and Methods), the finite size of the metasurface array (115 μm × 115 μm), and not perfectly collimated incident light. As expected, for a given value of $g$, the PA absorbance spectrum with smaller $\theta$ shows a reduced resonance FWHM, resulting in high $Q$ factors. The behavior of the $Q$ factors over the full parameter space has been investigated by calculating them from the absorbance spectra via a fitting approach using temporal coupled mode theory (TCMT) (for details see Materials and Methods). The highest $Q$ factor of nearly 100 can be obtained for $g$ = 600 nm and $\theta$ = 2° (**Figure 2d**). The $Q$ factors extracted from the experimental measurements follow the same trend but are lower than the simulated values (**Figure 2f**), which we again attribute to fabrication imperfections. Crucially, the maximally achievable $Q$ factors in our new BIC-driven PA platform are at least six times higher than that in the conventional solely gap-tuned systems (**Figure S3**).[40] From the $Q$ factor map, we can conclude that for fixed $g$, the $Q$ factor decreases with increasing $\theta$, which is mainly attributed to the increase of $\gamma_{rad, \theta}$. However, for smaller gaps ($g$ < 200 nm), the $Q$ factor does not monotonically decrease with an increase in $\theta$, especially at the UC regime, where $\gamma_{int}$ plays a dominant role, indicating the importance of the CC condition also for these $Q$ factors. This behavior is clearly demonstrated by the extracted $Q$ factors plots with fixed $g$ = 600 nm (580 nm) and 200 nm as a function of $\theta$ as shown in **Figure 2d** (simulations) and **Figure 2f** (experiments) and further supported by the absorbance spectra of PAs with $g$ = 200 nm and varied $\theta$ in **Figure S6**. Importantly, these results reinforce the versatility of our concept for tailoring absorption, near-fields, and $Q$ factors of the BIC governed PA resonances. High-resolution scanning electronic microscope (HR-SEM) images (**Figure 2g**) shows the fabricated unit cells of PA metasurface with varied $\theta$ used in above measurements (for details of PA metasurface fabrication see Materials and Methods).

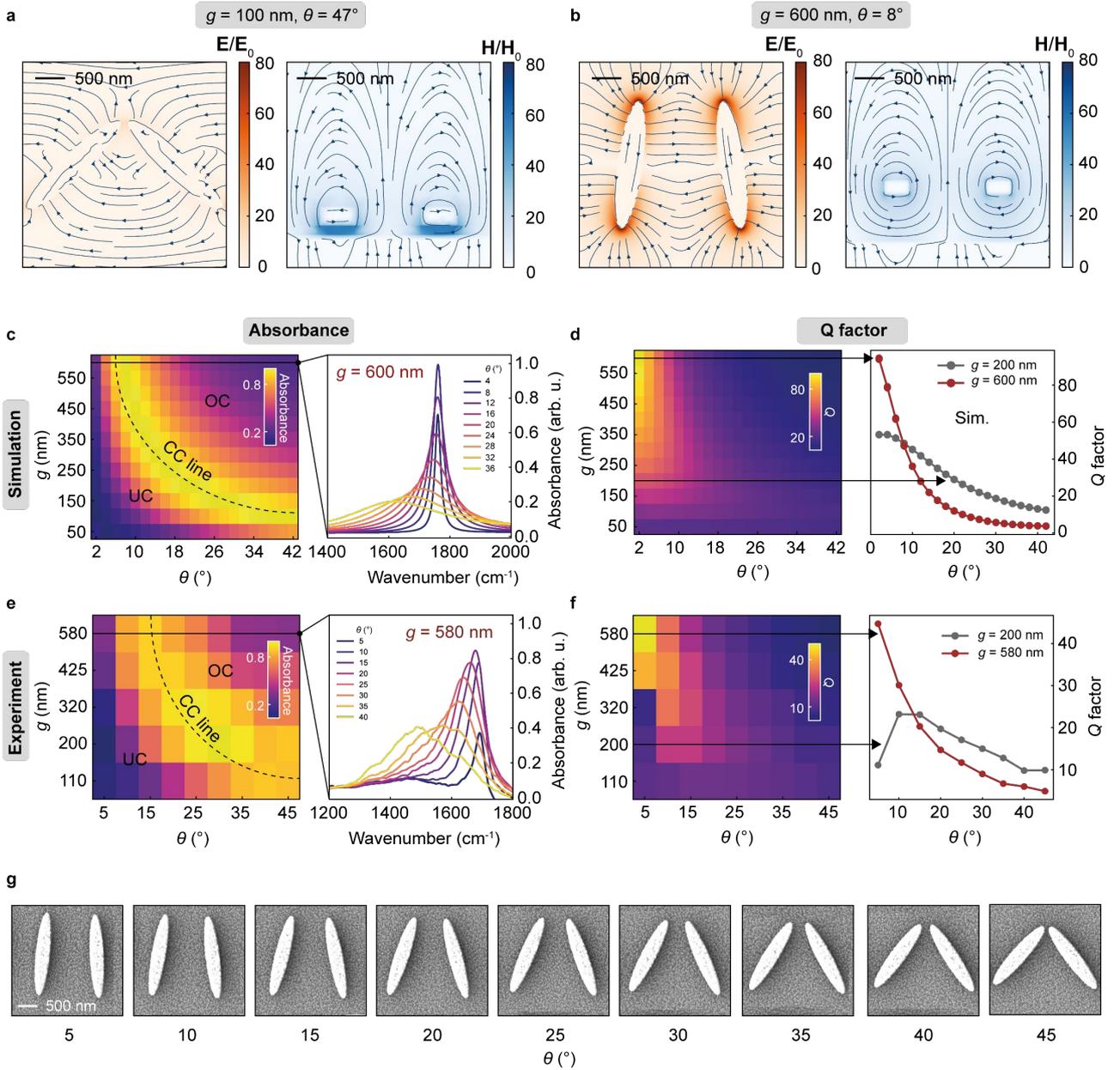

**Figure 2**. Tailored light-matter coupling in plasmonic BIC-driven PAs via simultaneous control over gap ($g$) and asymmetry parameter ($α$). (**a**) Near-fields of a PA with $g$ = 100 nm and $θ$ = 47 º at the CC condition. (**b**) Near fields of a PA with $g$ = 600 nm, and $θ$ = 8º at the CC condition. Orange (blue) colors indicated the **E** (**H**) field distribution in (a) and (b). (**c**) The 2D parameter space ($g$, $θ$) map for co-tuning the light-matter coupling regime in numerical simulation and extracted numerical absorbance spectra as a function of $θ$ with constant $g$ = 600 nm. (**d**) 2D map of calculated $Q$ factors from each absorbance spectra in (c) and extracted $Q$ factor plots as a function of $θ$ with constant $g$ (indicated by the grey and dark red lines). (**e**) Experimental realization of 2D ($g$, $θ$) map for co-tuning the absorbance coupling regime and extracted experimental absorbance spectra as a function of $θ$ with constant $g$ = 580 nm. (**f**) 2D Map of $Q$ factors obtained from the measured absorbance spectra in (e) and $Q$ factor plots with increasing $θ$ at the fixed $g$ (indicated by the grey

and dark red lines). (**g**) Top-view HR-SEM images that are used in panels (e) and (f) show unit cells with varying $\theta$ from 5° to 45°. Scale bar: 500 nm.

**Analytical coupling analysis and near-field enhancement evaluation**

The plasmonic BIC-driven PA system has been further analyzed using TCMT,[51] allowing us to separate the contributions of the intrinsic and radiative quality factors to the enhanced absorption (for details see Materials and Methods). For numerical analysis, we fixed the intrinsic damping rate ($\gamma_{int}$) for varying ellipse tilting angles $\theta$, because the BIC asymmetry factor solely affects the radiative damping rates,[52] thus the intrinsic $Q$ factor $Q_{int}$ only changes slightly with $\theta$, due to the shift of resonance positions, while $Q_{int}$ increases significantly with $g$ (**Figure 3a**). In contrast, the radiative damping rate $\gamma_{rad}$ strongly depends on both $g$ and $\theta$. From the map of radiative $Q$ factors (**Figure 3b**), $Q_{rad}$ decreases with increasing $g$ (for a fixed $\theta$), particularly with larger $\theta$, while for $\theta \lesssim 4°$, $Q_{rad}$ first increases and then drops with $g$ (**Figure S7**). Looking at fixed values of $g$ (indicated by the black arrows in **Figure 3b**), for large spacers ($g = 600$ nm), $Q_{rad}$ follows the characteristic BIC inverse quadratic dependency with respect to the asymmetry factor $\sin \theta$ ($Q_{rad} \sim \sin^{-2} \theta$),[44]. Decreasing the spacer thickness $g$ leads to a deviation from the ideal BIC regime, indicating strong near-field interaction between the plasmonic BIC and the induced mirror BIC (**Figure 3c**). Merging the intrinsic and radiative $Q$ factors to a total $Q$ factor (**Figure 3d**), reproduces panel (d) in Figure 2, which further corroborates the accuracy of our fitting approach. CC in the PA occurs for $Q_{int} = Q_{rad}$ and therefore we can visualize the different coupling regimes by mapping the difference $Q_{int} - Q_{rad}$ of the two $Q$ factors (**Figure 3e**). We find that the CC line reproduces the distinct behavior already observed in **Figure 2c**.

Importantly, at the CC line, both the $Q$ factors and the **E** field enhancements (noted as $|\mathbf{E}|^2/|\mathbf{E}_0|^2$) show an increase with increasing $g$ (see **Figures 2d**, **f** and **S7**), indicating that the **E** field enhancement is mostly governed by the BIC-like behavior of the metasurface layer, which can be tuned via $\theta$. On the other hand, from the map of $Q_{int}$ (**Figure 3a**), an increase of $Q_{int}$ with increasing $g$ can give rise to much stronger enhanced near fields at the CC regime. Therefore, in general, the highest $Q$ factors and largest near-fields enhancements can be obtained in a design with large $g$ and small $\theta$ (e.g., $g = 600$ nm, $\theta = 8°$). The maximum achievable near-field enhancements are above $10^4$ (**Figure S8**), while simultaneously providing strong electromagnetic hot-spots confined the outside of resonators, enabling substantial enhancement of light-matter interactions. Furthermore, the integrated near-field intensity enhancements confirm that the strongest enhancements indeed follow the CC condition (**Figure 3f**, for calculation details see Materials and Methods). The large near-fields obtained in a carefully tailored BIC-driven PA geometry are highly beneficial for the sensitive detection of surface-bound molecules, further strengthened by the freedom to flexibly tune the PA wavelength to overlap with the molecular fingerprints and the precise targeting of specific vibrational bands via the sharp resonances. Crucially, the accessible hot-spots and $Q$ factors of the resonances can be tailored simultaneously, providing

high spectral selectivity and sensitivity of analytes. Taken together, these features make our PA platform excellently suited to serve as molecular sensors.

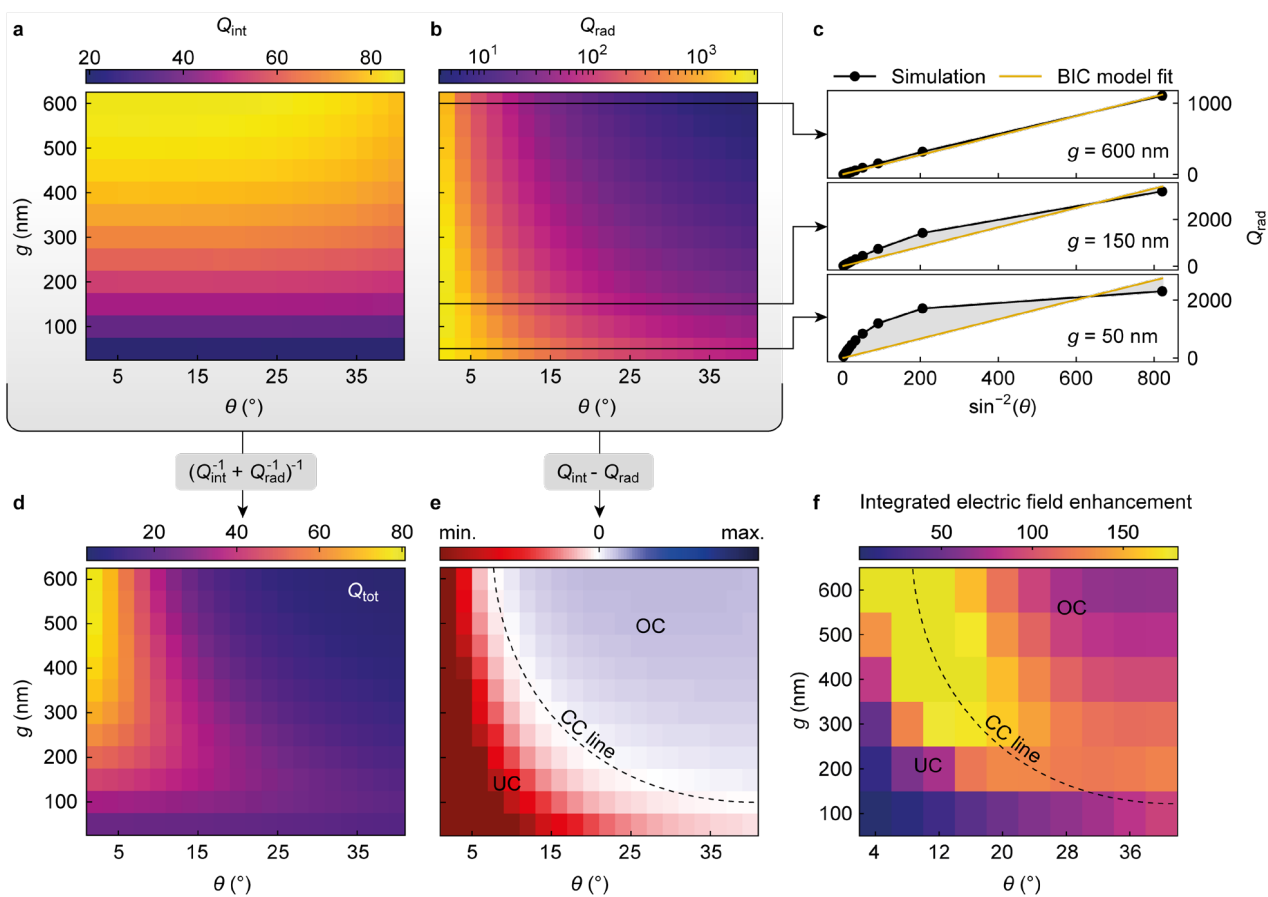

**Figure 3**. TCMT modelling and near-field enhancement evaluation of BIC-driven PAs. (**a**) $Q_{int}$ as a function of $g$ and $\theta$. (**b**) $Q_{rad}$ as a function of $g$ and $\theta$. (**c**) $Q_{rad}$ extracted from panel (b) as a function of the inverse square of the asymmetry parameter $\alpha^{-2}$ ($\alpha = \sin\theta$), fitted with a BIC model. For large gaps ($g$ = 600 nm), the resonance behaves like an ideal BIC, whereas smaller gap sizes lead to a deviation from the linear trend of the radiative $Q$ factor, indicating a strong near-field interaction of the plasmonic BIC with the mirror BIC. (**d**) Total $Q$ factor retrieved from panels (a) and (b). (**e**) Map of the difference between $Q_{int}$ and $Q_{rad}$ as a function of $g$ and $\theta$. (**e**) Integrated **E** field enhancement as a function of $g$ and $\theta$, where the resonance was placed approximately at 1650 cm$^{-1}$ ($S$ = 1.08).

## Pixelated BIC-driven PA metasurfaces for biomolecular detection

Biomolecules contain a rich fingerprint of vibrational absorption bands over a wide range of wavelengths in the mid-IR. To identify and quantify the trace amounts of molecular species, carefully engineered nanophotonic platforms capable of delivering resonances targeted at specific molecular signatures with high $Q$ factors and enhanced near fields for strongly enhanced light-matter interaction are essential. To highlight the molecular detection capability of our PA design, we evaluate the sensing performance of a PA realization with a large spacer thickness of $g = 600$ nm and varied $\theta$, covering the distinct UC ($\theta < \theta_{cc}$), CC ($\theta = \theta_{cc}$) and OC ($\theta > \theta_{cc}$) regimes. The value of $g = 600$ nm was chosen to simultaneously provide high $Q$ factors (i.e., high spectral selectivity) and strong near fields in accordance with the analysis presented in **Figure 3**. We first investigate different coupling regimes using PMMA coated metasurfaces (**Figure 4a**). **Figures 4b** shows the simulated and experimental absorbance spectra of the PA upon coating with an ultrathin layer of polymethylmethacrylate (PMMA) polymer. The thickness of the conformal PMMA layer is 2 nm in simulations and approximately 1.5 nm in experiments as confirmed by spectral ellipsometry measurements. Clearly, upon addition of an absorbing analyte layer, an increase in the intrinsic loss of the coupled system is expected, where $\gamma_{int} \rightarrow \gamma_{int} + \gamma_\mu$, and $\gamma_\mu$ is the loss of absorbing analyte. Therefore, the absorbance amplitude is further decreased by the addition of the analyte when the bare PA is in the UC regime, whereas we observe an absorbance increase in the OC regime. This behavior is attributed to an increase in $\gamma_{int}$, which drives the system closer to the CC point since $(\gamma_{int} + \gamma_\mu)/(\gamma_{rad}) > (\gamma_{int})/(\gamma_{rad})$. For the PA at the CC point, only small changes of the absorbance amplitude are expected, since the system is only shifted slightly away from the CC condition, which is reflected in both the numerical and experimental results. **Figure 4c** compares the change of absorbance amplitude ($\Delta A = A_{bare} - A_{sample}$) between the BIC-driven PA before and after analyte coating, clearly identifying the three coupling regimes based on the absorbance modulation.

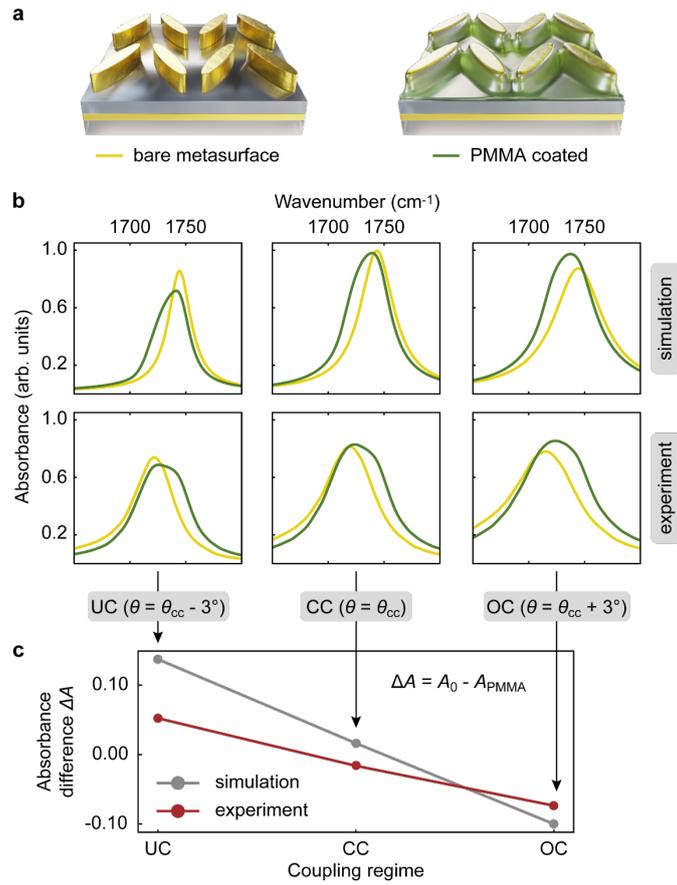

**Figure 4**. Coupling regimes in plasmonic BIC-driven PAs. (**a**) Illustrations of bare and PMMA coated metasurfaces. (**b**) Simulated and experimental spectra show the absorbance modulation of PAs in the UC, CC, and OC regime when coated with a thin (2 nm in simulation and 1.5 nm in experiment) layer of PMMA polymer. (**c**) Absorbance amplitude difference ($\Delta A$) for different coupling regimes upon the PMMA coating compared to the bare PA.

To demonstrate the capability of maintaining CC over a wide range of mid-IR wavelengths (1250 cm$^{-1}$ to 2000 cm$^{-1}$), we develop a pixelated metasurface of PAs at CC condition (**Figure 5a**), where each metapixel corresponds to a discrete frequency with near-unity absorbance (> 99.4 %) and high $Q$ factors (> 45) in simulations (**Figures S9** and **S10**). The full absorbance spectra of the PA pixels are shown in **Figure S9**. A broad wavelength coverage was achieved by linearly scaling the geometric dimensions by a factor $S$ and simultaneously tuning the ellipse tilting angle $\theta$ (**Figure 5a**). The versatility of tuning $\theta$ allows to achieve the CC condition of each pixel. To the best of our knowledge, this is the first report of a plasmonic BIC-driven PA concept, where one can easily achieve a broad range of resonances with near-unity absorbance, high $Q$ factors, strongly enhanced near fields, and precisely tailored coupling regimes, constituting a significant advance compared to conventional plasmonic PAs (lower $Q$ factor and limited resonance tuning)[33,38] and film-coupled low-loss resonators (reduced field confinement/enhancement and limited control over radiative decay rates).[24]

Since the highest analyte-induced absorbance modulation change and therefore the highest molecular sensitivity is expected in the UC regime (**Figure 4c**), we now experimentally realize a BIC-driven PA design for $\theta_{uc} = \theta_{cc} - 3°$. To demonstrate bimolecular sensing, a 2 nm thick conformal protein layer was applied on top of bare PA pixels. The absorbance spectra of bare pixelated PAs (grey color, denoted as $A_0$) and protein coated ones (blue color, $A_s$) are shown in **Figure 5b**. When the resonances of metapixels couple with the vibrational bands of the proteins (*i.e.*, amide I at ~ 1650 cm$^{-1}$ and amide II at ~ 1530 cm$^{-1}$), there is a pronounced decrease in absorbance amplitude. The IR absorbance spectrum of the protein can then be obtained from the envelopes of the respective absorbance spectra via $A$ = -log ($A_s/A_0$) (**Figure 5c**, grey curve), reproducing the characteristic amide bands. We have evaluated the sensing performance of the pixelated PA concept using a high-affinity biotin-streptavidin binding bioassay. Biotin molecules were attached to the Au surface using an established thiolation protocol (see details in Materials and Methods and **Figure S11**). The experimentally obtained vibrational absorbance spectra of streptavidin (**Figure 5c**, dark red curve) is referenced to the signal of thiolated biotin molecules (SH-biotin), where the SH-biotin molecules serve as capture sites to specifically bind the streptavidin. The measured amide I and II vibration bands agree well with the simulation, but show a reduced absorbance amplitude, which we attribute to impurities of the gold surface and the biotin monolayer. To study the performance of molecular sensing for all coupling regimes, pixelated PAs at the CC ($\theta = \theta_{cc}$), and the OC ($\theta = \theta_{cc} + 3$ °) conditions have also been realized and compared using the biotin-streptavidin bioassay, reproducing the behavior observed for the PMMA above (**Figure S11d**). Therefore, our BIC-driven PA concept shows strong promise for molecular identification over wide spectral ranges with high specificity and sensitivity, paving the way towards minimized and compact sensing platforms for biomedical assays and clinical diagnosis.

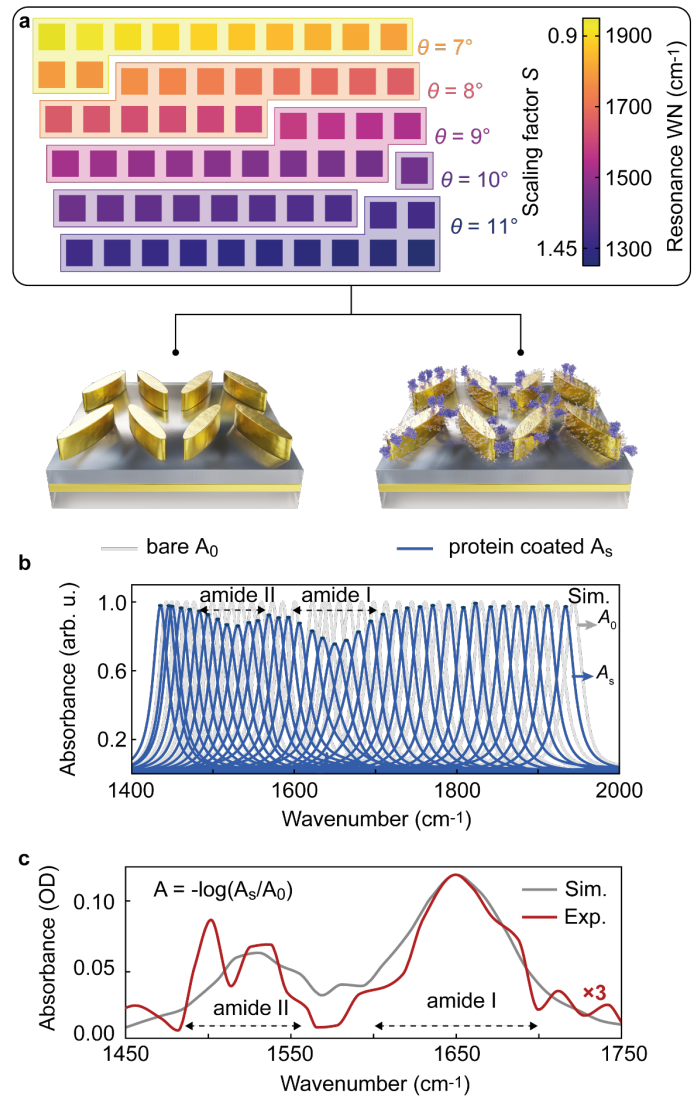

**Figure 5**. Pixelated plasmonic BIC-driven PAs for biomolecular sensing. (**a**) A pixelated BIC-driven PA metasurface realizes resonances over a broad mid-IR spectral range by simultaneously controlling $S$ and $\theta$ for a fixed gap size $g = 600$ nm. The corresponding $\theta$ of each metapixel is indicated for each metapixel and color shadowed boxes group pixels that share the same $\theta$. The scaling factor $S$ of the pixels increases linearly from the top left to the bottom right of the metasurface. (**b**) Simulated normalized absorbance spectra of the UC pixelated metasurface without molecular coating ($A_0$, grey color lines) and with a 2 nm thick protein coating ($A_s$, blue color). (**c**) Simulated and experimental protein absorbance spectra show the characteristic amide I (~ 1650 cm$^{-1}$) and amide II (~ 1530 cm$^{-1}$) band vibrations.

**Discussion**

We have numerically and experimentally demonstrated a new BIC-driven PA metasurface concept consisting of an array of tilted ellipse pairs in an MIM configuration. Our approach allows to precisely control the radiative loss channel in the PA system through both the asymmetric parameter ($\alpha$) and the insulator gap ($g$), delivering an additional degree of freedom for tailoring light-matter coupling. Significantly, such a BIC-driven PA enables multiple tuning knobs to precisely tune the coupling regimes through the in-plane ellipse tilting angle $\theta$ at any constant gap, which allows selective strong enhancement of either the electric or the magnetic fields. Crucially, we have found that the $\theta$-induced radiative loss mainly accounts for the electric field enhancements, while the $g$-induced radiative loss is responsible for controlling the formation of the magnetic mode, enabling flexible resonant systems for diverse applications. Furthermore, resonances tuned to the CC regime over a broad range of mid-IR wavelengths can be achieved by both linear scaling the geometric parameters by a factor of $S$ and tuning the tilting angle $\theta$ correspondingly, while maintaining high $Q$ factors and near-unity absorbance (> 99.4 %) throughout. Our results demonstrate that plasmonic BIC-driven PAs can deliver efficient light-matter coupling with the highest $Q$ factors (~ 100) and strongest near-field enhancements (~ $10^4$) found for small values of $\theta$ and large values of $g$, for example, $\theta = 8$ ° and $g = 600$ nm, outperforming traditional PA systems. We have further demonstrated a pixelated PA metasurface with resonances in a range of 1200 cm$^{-1}$ to 2000 cm$^{-1}$, in different coupling regimes (UC, CC and OC) and have applied the concept for biomolecular detection with high sensitivity and spectral selectivity. This BIC-driven PA concept opens a new direction towards miniaturized sensing platforms for on-chip and in-situ detection. In the future, a BIC-driven optofluidic PA, where the analyte simultaneously acts as spacer layer and analyte channel, which maximizes the utilization of the near-field hotspots may deliver greatly enhanced sensitivities for molecular trace detection in biomedical diagnostics or environmental monitoring.

**Materials and Methods**

**Numerical simulation**

All numerical simulations have been performed using the finite element solver contained in CST Microwave Studio 2021 (Dassault Systèmes), where periodic boundary conditions are applied and the incident light ($k$) is perpendicular to the metasurface plane with $x$-polarization (TM). Tabulated permittivity values of Au, PMMA and protein are taken from the literature[53–55]. The refractive indices of silica glass and CaF$_2$ were assumed to be lossless with $n_{SiO2} = 1.5$ and $n_{CaF2} = 1.399$, respectively. Tuning of the PA resonances throughout the mid-IR spectral range is implemented by applying a scaling factor $S$ to the geometric parameters ($P_{x,0}$, $P_{y,0}$, $A_0$ and $B_0$) of the unit cell (**Figure S1**). Throughout the work, a constant Au resonator height of $h = 200$ nm is used. Besides, the conventional gap-tuned PAs consisting of brick, disk, symmetric paired ellipse arrays have been simulated (see SI with more parameter details). The thickness of the conformal absorbing analyte layer is set to 2 nm in simulations. The integrated electric near-fields enhancement has been evaluated by the ratio between the integrated enhanced electric fields over a certain volume and the $|\mathbf{E}_0|^2$ over the same volume. This volume was defined by the periodicity in $x$, $y$ direction and a height from the Au ground plane up to 100 nm above the top surface of the Au resonator ($H = h + g + 100$ nm).

**Fabrication of BIC-driven PA metasurfaces**

Starting from a commercial silica glass substrate, a 200 nm thick layer of Au was first evaporated as a mirror ground plane, followed by the evaporation of a dielectric $CaF_2$ layer with varying thickness in a range of 50 nm to 600 nm. BIC-driven metasurface structures were fabricated on top of the layer stack by a standard electron-beam lithography process (30 kV voltage, 15 μm aperture) using a double layer of positive PMMA resist (950 K, A4, Microresist).[10] Afterwards, an adhesion layer of 5 nm Ti and 200 nm of Au were deposited by electron-beam assisted evaporation, and the final metasurface structures were obtained by wet-chemical lift-off process (remover 1165, Microresist).

**Optical characterization**

To characterize the optical responses of the fabricated PA metasurfaces, a mid-IR spectral imaging microscope (Spero from Daylight Solutions Inc., USA) has been used for all measurements in this work. A low magnification objective (4 ×, NA = 0.1 with 2 mm² field of view) was used to ensure that large pixelated metasurfaces can be imaged simultaneously. The Spero microscope is equipped with four tunable quantum cascade lasers continuously covering the mid-IR range from 948 cm$^{-1}$ to 1800 cm$^{-1}$. A laser scanning step size of 2 cm$^{-1}$ was used. All optical measurements were conducted in reflection mode and normalized to the reflection signal of a plain Au mirror (thickness 200 nm). To eliminate backscattering effects, the reflectance signal from an unpatterned area on the metasurface chip was collected as a reference. The final spatially resolved metasurface reflectance signals were then obtained by subtracting the reference data from the sample data. Metapixel spectra were obtained by averaging the hyperspectral reflectance data over the corresponding image pixels.

**Analytical analysis**

To extract the intrinsic ($\gamma_{int}$) and radiative ($\gamma_{rad}$) damping rates of PA resonances with wavenumber $\nu_0$, we utilized TCMT[56] to describe a resonator with a single port supporting reflected waves and a single resonance coupling to the far-field mediated by the coupling constant $\kappa = \sqrt{2\gamma_{rad}}$ and an intrinsic loss channel, damping the resonance with a rate $\gamma_{int}$. In this system the absorbance ($A$) is given by the equation

$$A = 1 - R = \frac{4\,\gamma_{rad}\,\gamma_{int}}{(\nu - \nu_0)^2 + (\gamma_{rad} + \gamma_{int})^2}$$

which can be directly fitted to our simulated absorbance spectra. The radiative, intrinsic and total $Q$ factors can then be calculated as $Q_{rad} = \frac{\nu_0}{2\gamma_{rad}}$, $Q_{int} = \frac{\nu_0}{2\gamma_{int}}$ and $Q_{tot} = \frac{\nu_0}{2(\gamma_{rad}+\gamma_{int})}$, respectively. To analyze the damping rates for different values of $g$ and $\theta$, we use the fact that $\gamma_{rad}$ and $\gamma_{int}$ are independent variables, i.e. material intrinsic losses have no effect on the radiative damping rate, by fitting absorbance spectra with different $\theta$ at a given $g$ simultaneously with a shared value of $\gamma_{int}$, which adds stability to the fitting procedure.

**Molecular sensing**

For the PMMA coated PA, a layer of 0.1 % PMMA (950 K, diluted in anisole) was spin-coated onto the bare PA (3000 rpm, 1 min), followed by a 5 min baking at 180 ºC. Ellipsometric measurements (HS-190, J.A. Woollam VASE) on a clean silicon substrate coated with the same layer were used to precisely determine the thickness of the PMMA film.

For the biomolecular detection experiments, a layer of capture SH-biotin molecules (HS-$(CH_2$-$CH_2$-$O)_5$-$OCH_2CH_2$-NH-biotin, Prochimia Surfaces, Poland) was immobilized on the freshly prepared PA metasurfaces. The applied concentration of SH-biotin solution is 1.2 mg·mL$^{-1}$ in a degassed mixture of deionized water and ethanol with volume ratio of 1:1. After incubation of the metasurfaces in SH-biotin solution for 4 hours, the metasurface was rinsed continuously by ethanol and dried by a stream of $N_2$ flow. Then, the functionalized metasurface was incubated in streptavidin solution for 2 hours by drop casting with a volume of 80 μL at room temperature, where the used concentration of streptavidin was 16 μM. After incubation, the streptavidin immobilized metasurface was rinsed by phosphate-buffered saline (PBS) buffer (1×) and dried by $N_2$ flow. After each immobilization step, the PA mestasurface was transferred to the Spero mid-IR microscope for spectroscopic imaging.

**Acknowledgements**

This work was funded by the Deutsche Forschungsgemeinschaft (DFG, German Research Foundation) under grant numbers EXC 2089/1 – 390776260 (Germany's Excellence Strategy) and TI 1063/1 (Emmy Noether Program), the Bavarian program Solar Energies Go Hybrid (SolTech), and the Center for NanoScience (CeNS). S.A.M. additionally acknowledges the Lee-Lucas Chair in Physics and the EPSRC (EP/W017075/1).

**Author Contributions**

J. W proposed the idea, performed the numerical simulation and experimental work, and drafted the manuscript. T. W performed the analytical TCMT model analysis and data processing. A. A., T. W., S. M. and A. T. contributed to the data discussion and the manuscript revision. A. T. initiated and supervised the project. All the authors completed the writing of the manuscript.

**Conflict of Interest**

The authors declare no competing financial interest.

## Supporting Information

## Mirror-coupled plasmonic bound states in the continuum for tunable perfect absorption

Juan Wang[1], Thomas Weber[1], Andreas Aigner[1], Stefan A. Maier[3,2,1] and Andreas Tittl[1]*

[1]Chair in Hybrid Nanosystems, Nanoinstitute Munich, Faculty of Physics, Ludwig-Maximilians-Universität München, 80539 München, Germany

[2]The Blackett Laboratory, Department of Physics, Imperial College London, London SW7 2AZ, United Kingdom

[3]School of Physics and Astronomy, Monash University, Clayton, Victoria 3800, Australia

*e-mail: andreas.tittl@physik.uni-muenchen.de

**Plasmonic BIC-driven PA design**

The proposed plasmonic BIC-driven PA consists of periodically arranged paired tilting ellipses. Here the ellipse tilting angle $\theta$ defines the asymmetric properties of the consituent metatoms (**Figure S1**). Through the whole work, we used optimized geometric parameters of a unit cell including periodicity in $x$ and $y$ directions ($P_{x,0} = P_{y,0} = 3200$ nm), long and short axis of ellipses ($A_0 = 2000$ nm, $B_0 = 400$ nm), and height of the ellipse ($h = 200$ nm). A universal scaling approach is used to extend the resonance to other spectral wavelengths, where $P_x = S \times P_{x,0}$, $P_y = S \times P_{y,0}$, $A = S \times A_0$, $B = S \times B_0$.

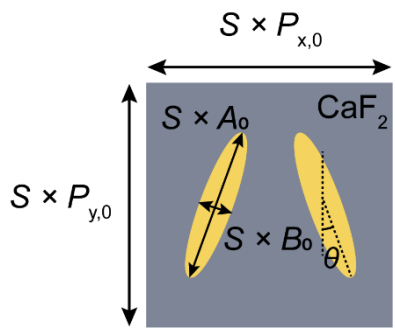

**Figure S1**. The top-view shematic of a unit cell showing the used parameters in this plasmonic BIC-driven PA.

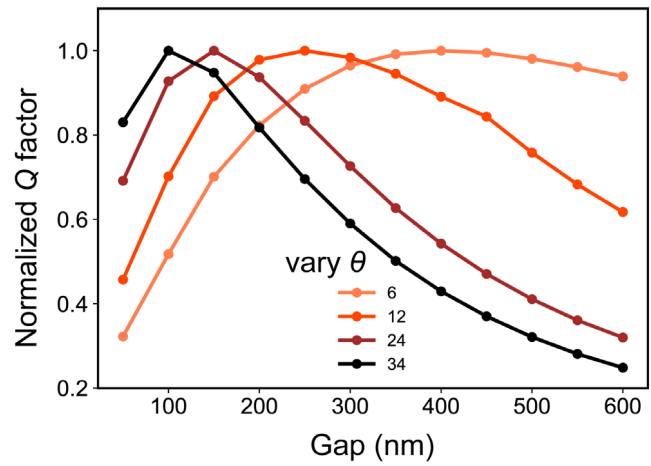

**Figure S2**. The normalized $Q$ factors as a function of gap at different tilting angle $\theta$ analyzied by the TCMT model.

**Conventional gap-tuned mettalic PAs**

The conventional gap-tuned metallic PAs consisited of disk and brick arrays have been investigated by the numerical simulation (**Figure S3**). It is found that the absorbance spectra of both PAs consitsed of disks and bricks show similar tendency with increasing $g$, which is also similar to that we have observed in symmetry-broken protected BIC-driven PAs with fixed asymmetric parameter $\alpha$ (**Figure 1c**). The extracted $Q$ factors from both solely gap-tuned PAs are all below 18, which is six times lower than that of the plasmonic BIC-driven PAs.

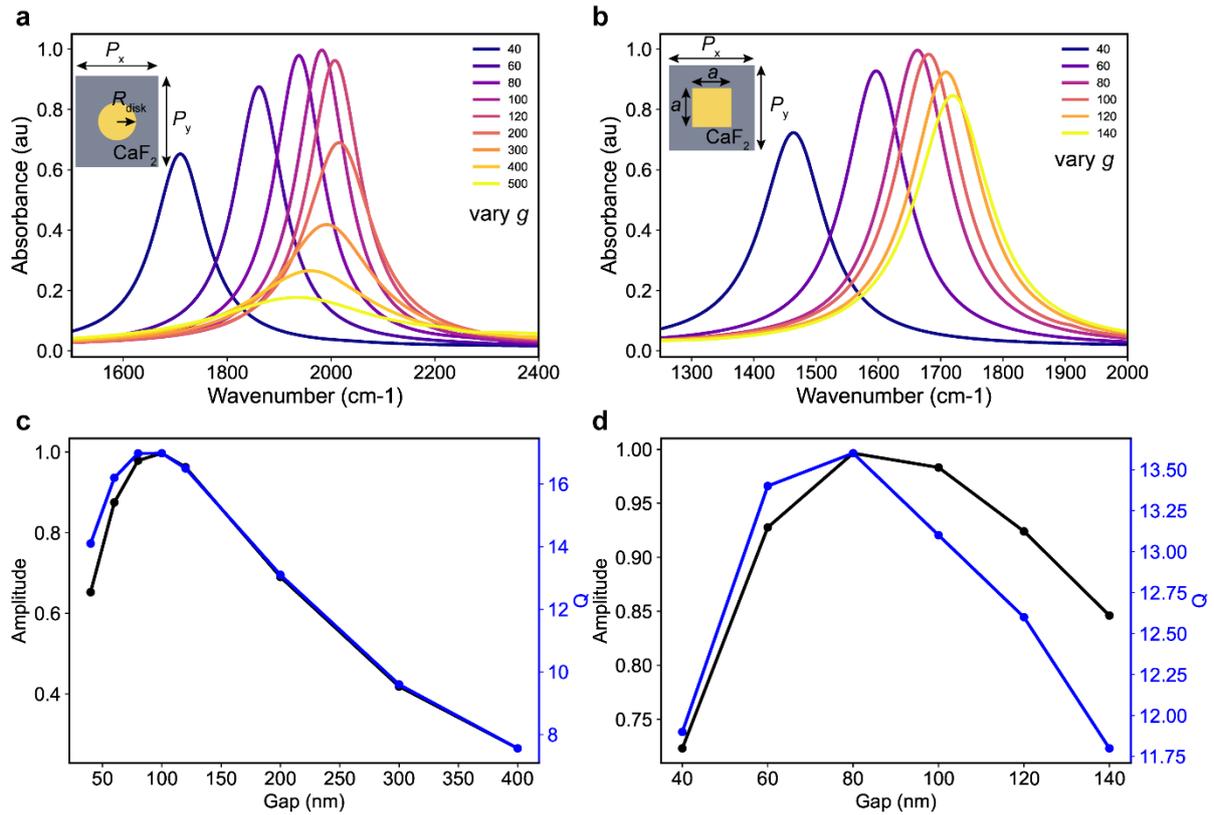

**Figure S3**. The conventional gap-tuned mettalic PAs. The gap-tuned PAs consisted of an array of disks (**a**) and bricks (**b**). (**c**) The maximum absorbance intensity and correspondingly calcuated $Q$ factors of PAs consisted of disk arrays, where $P_x$ = 3600 nm, $P_y$ = 4000 nm, $R_{disk}$ = 800 nm (radius of disk), $h_{disk}$ = 100 nm (height of disk). (**d**) The maximum absorbance intensity and extracted $Q$ factors of PAs consisted of brick arrays, where $P_x$ = $P_y$ = 3600 nm, $a$ = 1600 nm, $h_{brick}$ = 150 nm (height of brick). The top-view schematic are shown in inset.

**Metallic PAs consisted of symmetric paired ellipses**

Compared to the plasmonic BIC-driven PAs with asymmetric paired tilting ellipses, the resonances of PAs with symmetric ($\theta$ = 0 °) unit cells disapper, regardness of the gap size (**Figure S4a**). But it is worth noting that if we re-optimize the parameters of PAs composed of peridic symmetric paired ellipse arrays, there shows new resonant peaks and these resonances can be well-tailored by the gap (**Figures S4b** and **c**), where the radiative loss only tuned by the gap and is analogy to the gap-tuned system.

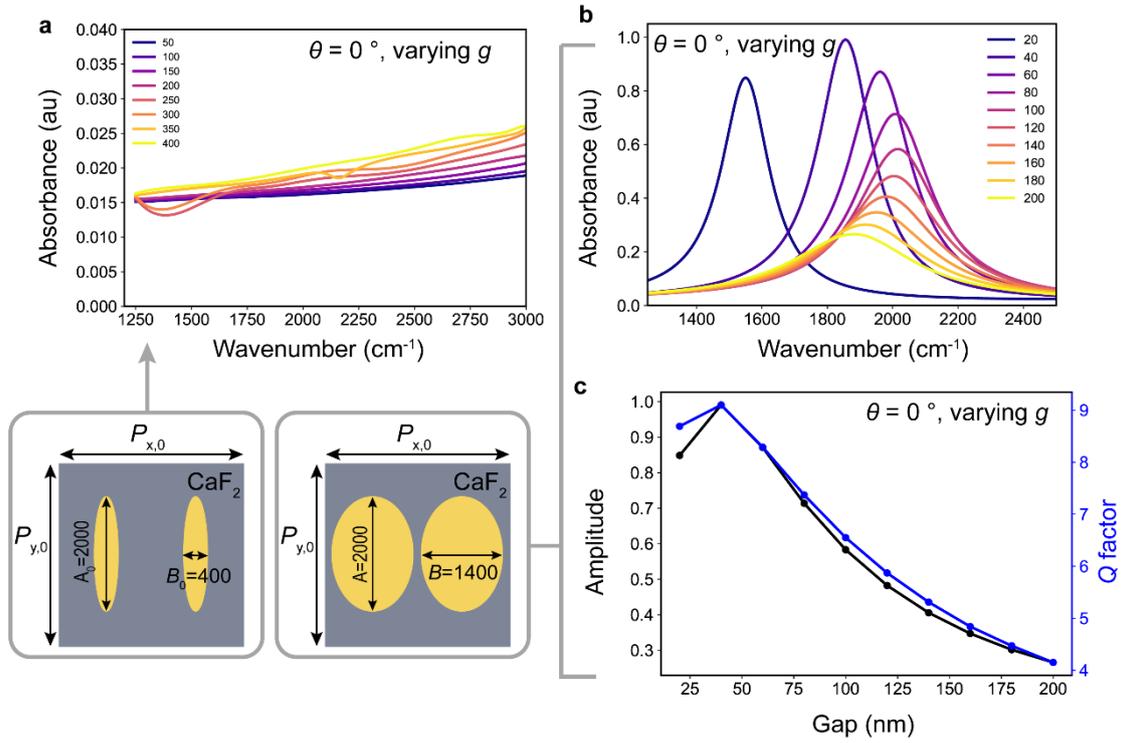

**Figure S4**. The mettalic PAs consisted of symmetric paired ellipses. (**a**) The simulated absorbance spectra of PAs consisted of symmetric paired ellipses ($\theta = 0$ º) with varied $g$. (**b**) Simulated absorbance spectra of re-optimized only gap-tuned PAs composed of symmetric ellipses with different $A$ ($A = 2000$ nm) and $B$ ($B=1400$ nm) as a function of $g$, where $P_{x,0} = P_{y,0} = 3200$ nm, $h = 200$ nm. (**c**) The corresponding ampiltude of absorbance and extacted $Q$ factors as a function of $g$ from (b). Here, all $S = 1$.

**Near-fields of BIC-driven PA at the UC and OC regime**

To compare the near-fields at the CC regimes with that at the UC and OC regime, we have further explored the near-fields distribution of BIC-driven PAs with both smaller $g$ and $\theta$ (**Figures S5a** and **b**), and larger $g$ and $\theta$ (**Figures S5c** and **d**). Obviously, at the UC regime, the $\gamma_{int}$ plays a dominant role, the field enhancement is much weaker compared to that at the CC regime, but they shows the same field vector distribution pattern. In contrast, for the PA at the OC regime, as expected, due to the large $\gamma_{rad}$, the near-field enhancements are also much weakened.

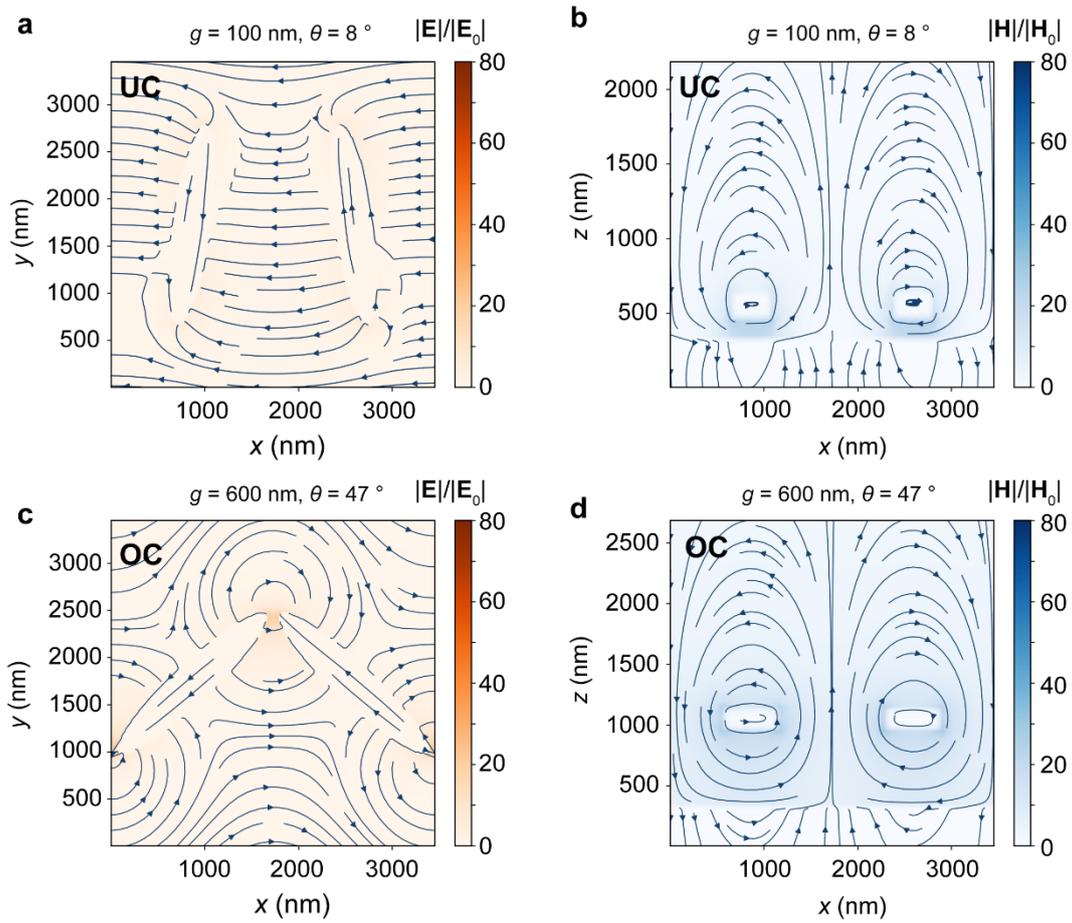

**Figure S5**. The near-fields map of BIC-driven PAs (**a**) at the UC regime with $g$ = 100 nm, $\theta$ = 8 °, and (**b**) at the OC regime with $g$ = 600 nm, $\theta$ = 47 °. The **E** (**H**) fields are indicated by the orange (blue) color.

**Absorbance spectra of BIC-driven PAs with varied $\theta$ at $g$ = 200 nm**

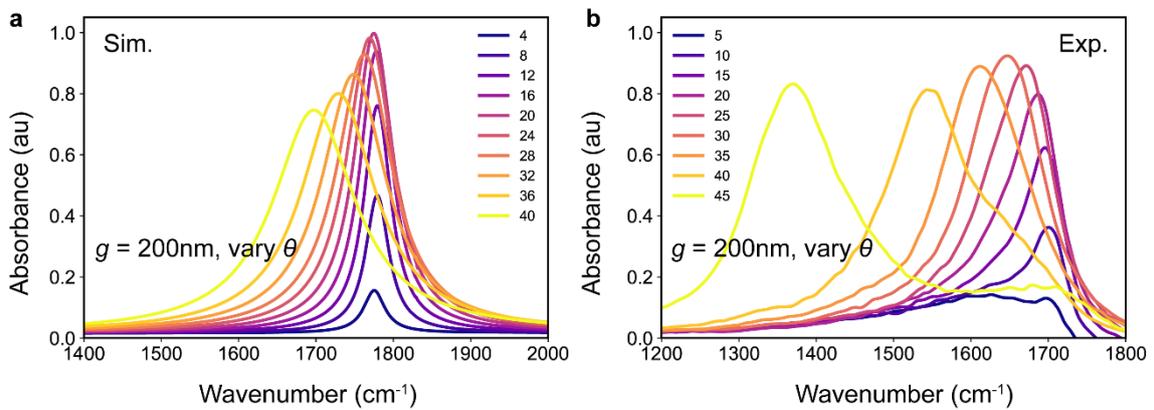

**Figure S6**. The exampled absorbance spectra plots of plasmonic BIC-driven PAs as a function of $\theta$ with constant $g = 200$ nm obtained by (**a**) numerical simulations and (**b**) experimental measurements.

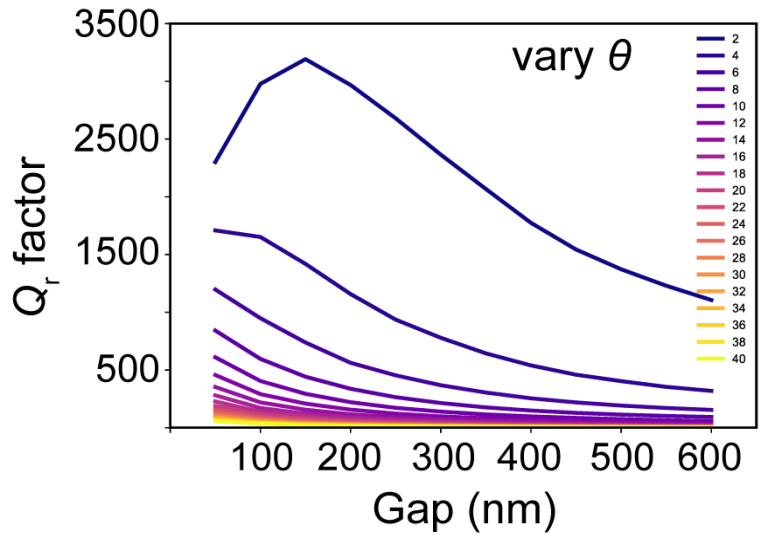

**Figure S7**. The radiative factor ($Q_r$) as a function of gap ($g$) with different $\theta$ from the TCMT analysis.

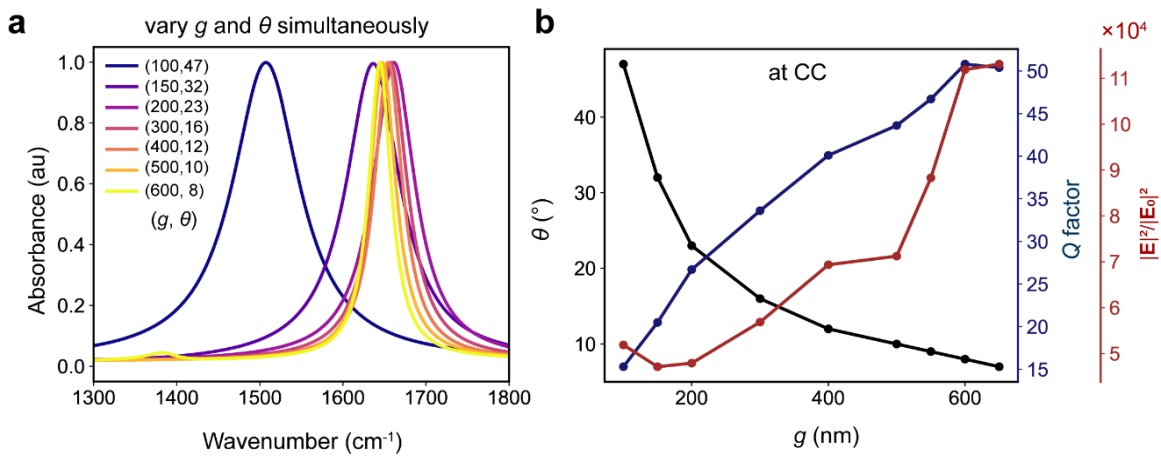

**Figure S8**. $g$ and $\theta$ simultaneously tuned plasmonic BIC-driven PAs. (**a**) The absorbance spectra of PA at the CC condition with desired $g$ and $\theta$. (**b**) The corresponding $\theta$, $Q$ factor and the magnitude of electric fields enhancements as a function of $g$ at their own CC condition.

## Pixelated BIC-driven PA metasurfaces design

The BIC-driven PAs enable us to tailor the resonances over a wide spectral range with both high $Q$ factors and near-unity absorbance via the in-plane asymmetric tilting angle $\theta$ and a universal scaling appracoh by a factor of $S$ on the optimized parameters of a unit cell at the fixed $g$. Therefore, we can flexibly achieve a broad range of ultrasharp resonances as exampled in **Figure S9** with $g$ = 600 nm, but varying $S$ and $\theta$ for each metapixel.

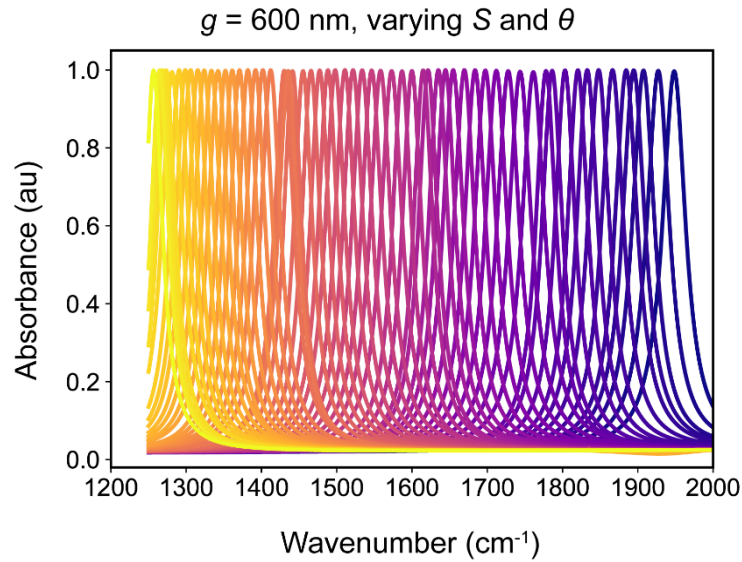

**Figure S9**. The simulated absorbance spectra of pixelated BIC-driven PAs with varying $S$ factor and $\theta$ to extend the resonances over a wide mid-IR spectral range. Here, $g$ = 600 nm and $S$ is in a range of 0.9 – 1.45 with 60 steps.

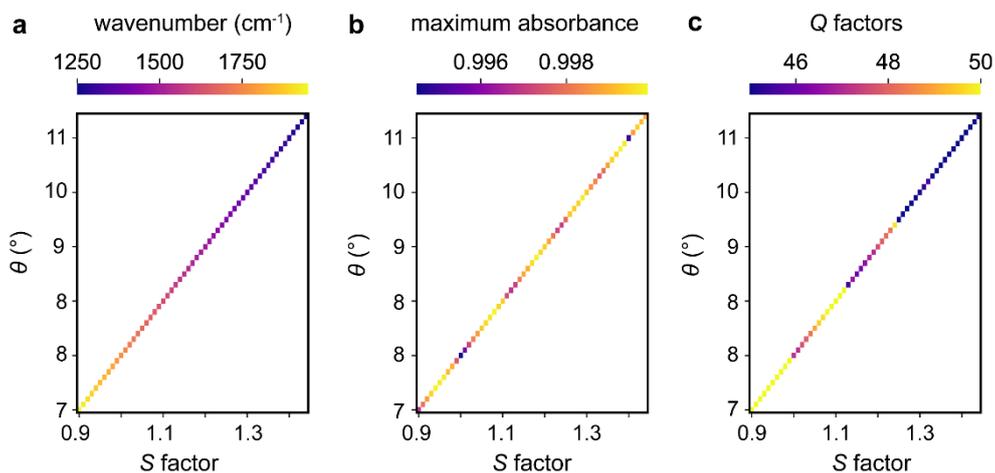

**Figure S10**. The programmed pixelated BIC-driven PAs with a wide range of mid-IR resonances by simultaneously tuning both the $S$ factor and $\theta$. (**a**) A universal method used to scale the resonances by a linear increase in $S$ factor with achieved resonances from 1250 to 2000 cm$^{-1}$. (**b**) The $\theta$ is tailored accordingly with the $S$ factor to reach the CC condition with near-unity absorbance for each metapixel. (**c**) The realized

resonances with high *Q* facors are all above 45 as indicated by the color bar. Here, the $g$ = 600 nm is used and keep it constant for all metapixels in this PA meatsurface.

**Pixelated plasmonic BIC-driven PA metasurfaces for biomolecular sensing**

After each step of surface functionzation of pixelated BIC-driven PA metasurfaces, they have been tranferred to the Spero mid-IR microscope to record the spectroscopic imaging. Through a custom Python code, the absorbance spectra of each metapixel can be extracted (**Figures S11a** and **b**). To clearly highlight the absorbance amplitude change after coating a layer of analyte on the metasurfaces, the initial absorbance spectra ($A_0$) are all normalized (grey curves, **Figures S11a** and **b**). Then the absorbance spectra of target analyte coated metasurfaces ($A_s$, orange curves, **Figures S9a** and **b**) are refered to the initially normalized spectra. **Figure S11c** shows one of selected absorbance spectra from **Figure S11b** (indicated by the pink star) to indicate the amplitude change upon the strong coupling between the resonant metapixels and the vibration bands of the immobilized analyte on top of metasurfaces. Afterwards, an equation of $A = -\log(A_s/A_0)$ was applied to obtain the molecular vibration band absorbance spectra (**Figure S11d**), indicating that the pixelated PAs at the UC regime with a positive IR absorbance peak, in conrast to the OC regime with a negtive IR absorbance peak of coated analytes. This confirms an excellent agreements with our theory.

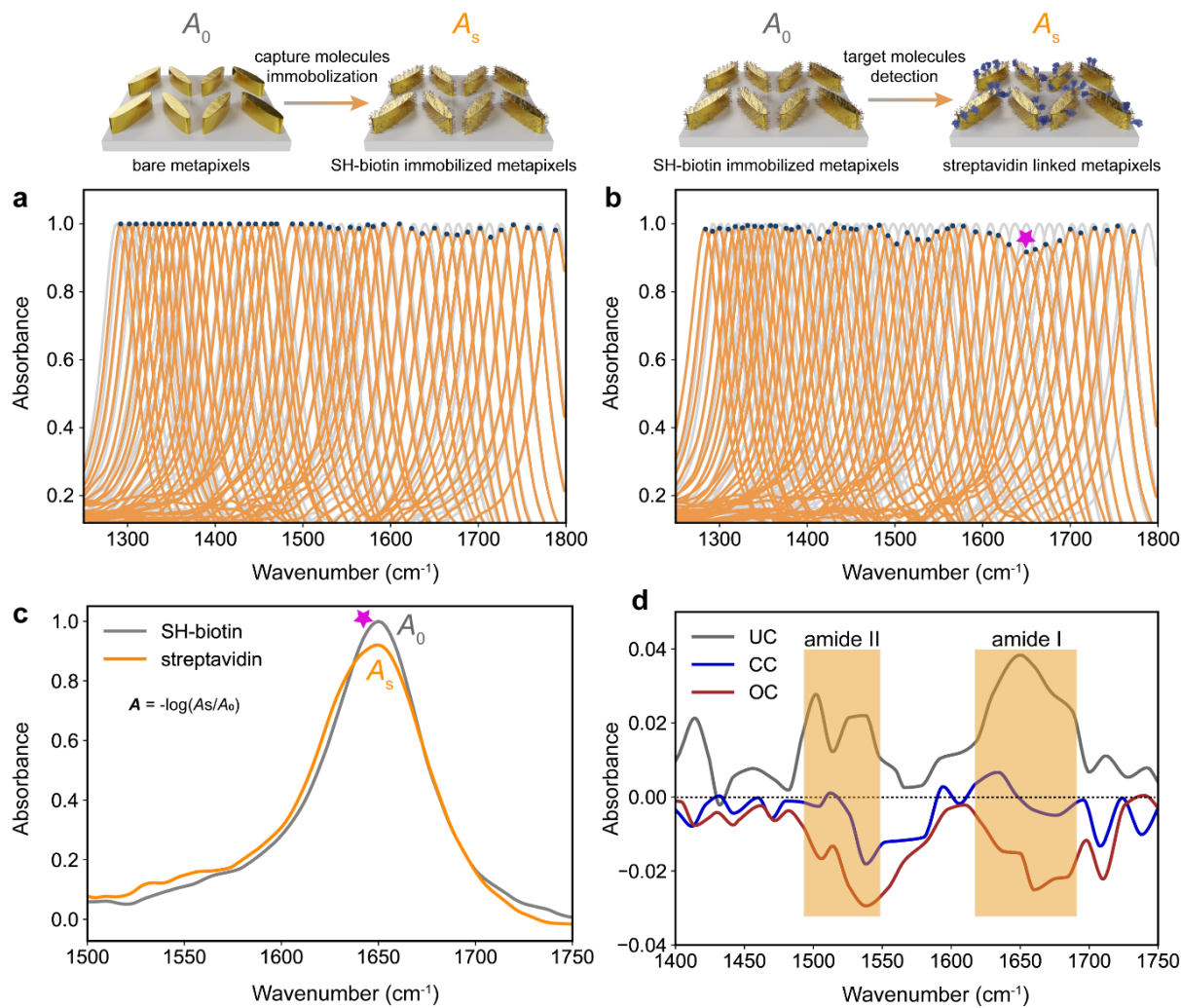

**Figure S11**. Experimental measurements of pixelated plasmonic BIC-driven PAs for protein sensing. (**a**) The absorbance spectra of metapixels without molecular coating (grey colour) and with capture molecules functionalization (yellow colour). (**b**) The absorbance spectra of capture molecules immobilized metasuraces (grey colour) and the subsequent streptavidin linked metasurfaces (yellow colour). All absorbance spectra indicated by the yellow color were referred to the previously treated metasurfaces (indicated by the grey colour). (**c**) The selected absorbance spectra from (b), indicating the obvious absorbance modulation change upon the coupling of resonant metapixels with analyte vibration bands. (**d**) The streptavidin absorbance spectra indicating the characteristic vibrational bands of amide I and II, where the plasmonic BIC-driven PAs at the UC, CC and OC regimes are used, respectively.